\documentclass[acmlarge,nonacm]{acmart}
\usepackage[utf8]{inputenc}
\usepackage[T1]{fontenc}

\usepackage{booktabs}
\usepackage{subfigure}
\usepackage{amsmath}
\usepackage{pmboxdraw}
\usepackage{multirow}
\usepackage{tabularx}
\usepackage{adjustbox}
\usepackage{multicol}
\usepackage{pmboxdraw}
\usepackage{icomma}

\usepackage{phdmisc}

\usepackage{graphicx}
\graphicspath{{figures/}}

\begin{document}

\title{Hypergraph-of-Entity}
\subtitle{A General Model for Entity-Oriented Search}

\author{José Devezas}
\email{jld@fe.up.pt}
\orcid{0000-0003-2780-2719}

\author{Sérgio Nunes}
\email{ssn@fe.up.pt}
\orcid{0000-0002-2693-988X}

\affiliation{%
  \institution{INESC TEC \& Faculty of Engineering, University of Porto}
  \streetaddress{Rua Dr. Roberto Frias, s/n}
  \postcode{4200-465}
  \city{Porto}
  \country{Portugal}
}

\begin{abstract}
  The hypergraph-of-entity was conceptually proposed as a general model for entity-oriented search. However, only the performance for ad hoc document retrieval had been assessed. We continue this line of research by also evaluating ad hoc entity retrieval, and entity list completion. We also attempt to scale the model, so that it can support the complete INEX 2009 Wikipedia collection. We do this by indexing the top keywords for each document, reducing complexity by partially lowering the number of nodes and, indirectly, the number of hyperedges linking terms to entities. This enables us to compare the effectiveness of the hypergraph-of-entity with the results obtained by the participants of the INEX tracks for the considered tasks. We find this to be a viable model that is, to our knowledge, the first attempt at a generalization in information retrieval, in particular by supporting a universal ranking function for multiple entity-oriented search tasks.
\end{abstract}


\maketitle

\section{Introduction}
\label{sec:intr}

Information retrieval includes a wide range of tasks that frequently depend on, or at the very least benefit from, cross-referencing information locked within heterogeneous data sources. In entity-oriented search, there is frequently a combination of corpora and knowledge bases, and a strong reliance on the integration of unstructured and structured data. Perhaps the most straightforward approach to tackle this problem is to store text in an inverted index and entities and their relations in a triplestore, and then separately compute and combine signals from each storage unit~\citep{Bast2016,Bast2012b,Lashkari2017}. Despite the importance of the cross-referencing aspect, few attention has been given to the relations within and across corpora and knowledge bases, in the design of representation models. Joint representation models have been explored through the usage of the inverted index, to store virtual documents built from passages mentioning the indexed entities~\citep{Bautin2007,Dietz2015} and, while \emph{mention} relations were implicitly captured in the process, no other relations (e.g., \emph{entity} relations) are available in the model to inform retrieval. Representation learning has also been used to find a common word and entity embedding space~\citep{Gysel2016} and, despite capturing latent relations between words and entities, these are not explicit or particularly exploited for retrieval. On the other hand, graph-based models are focused on the explicit representation of relations, be it intra-document, among terms~\citep{Blanco2012,Rousseau2013} or capturing syntactic and semantic dependencies~\citep{Widdows2002,Cancho2005}, inter-document, based on any type of links between documents~\citep{Broder2000}, or even based on document-entity relations, resulting from the annotation of entity mentions that point to their instance in a knowledge base~\citep{Guo2011}. These are some of the reasons that make graphs viable to support a retrieval process based on the cross-referencing of information locked within text with information directly expressed as triples. The challenge is how to build such a model in a way that it is as complete and useful as possible, while remaining efficient.

The hy\-per\-graph-of-entity~\citep{Devezas2019} was proposed as a joint representation model for text, entities and their relations, with the random walk score as a universal ranking function. This retrieval model provides a starting point not only for exploring the computation of multiple tasks from entity-oriented search, but also for finding answers based on a common source of cross-referenceable information. The authors evaluated the task of ad hoc document retrieval, proposing an analogous method for solving ad hoc entity retrieval, while also indicating related entity finding and entity list completion as part of four tasks that could be generalized using their model. In this work, we explore the potential of the hypergraph-of-entity for ad hoc entity retrieval, as well as for entity list completion, continuing with the generalization line that was proposed.

Out of the four proposed tasks, we do not directly experiment with related entity finding, for two reasons. First, there is no dataset combining a corpus with a knowledge base that provides topics and relevance judgments for the four tasks --- we use the INEX 2009 Wikipedia collection, which provides relevance judgments for the ad hoc document retrieval, ad hoc entity retrieval, and entity list completion. Secondly, we argue that entity list completion is a generalization of related entity finding, where the latter only takes one entity as input, while the former takes one entity, as well as examples of related entities, that work as relevance feedback. Both related entity finding, and entity list completion are defined as follows. Given an entity, a target type, and a relation, find other entities of the given target type that respect the specified relation. The difference between the two tasks lies in the fact that entity list completion also includes example entities to drive results towards similar entities. We opted, however, to simplify the view on these tasks, defining entity list completion as the task of finding other similar entities, given a set of input entities. For the particular case when this set has unitary cardinality, we consider it to be the same as related entity finding. This simplified definition makes particular sense in the context of hypergraph-of-entity, as the model does not store entity or relation types, making any type restrictions useless.

In order to be able to index the complete INEX 2009 Wikipedia collection, we are also required to reduce the size of the hypergraph-of-entity. In order to do this, and since we do not control the number of documents in a collection, we retain only representative keywords for each document. This not only reduces the quantity of nodes --- and, indirectly, the quantity of hyperedges --- but it also improves the overall quality of the model and its retrieval effectiveness. Accordingly, the contribution of this work is two-fold:

\begin{itemize}
  \item We tackle the performance issues of the hypergraph-of-entity by reducing the number of terms in each document through keyword extraction;

  \item We assess the performance of a universal ranking function for three entity-oriented search tasks:
  \begin{itemize}
    \item Ad hoc document retrieval;
    \item Ad hoc entity retrieval;
    \item Entity list completion.
  \end{itemize}
\end{itemize}

We rely on a simplified version of TextRank for keyword extraction. This version is more efficient but less effective than the original, which still results in an adequate amount of text usable for indexing. Different tasks are not comparable among each other. They are only comparable with their corresponding baselines. Nevertheless, our main goal is to understand whether performance is consistent and acceptable. Thus, our experiments are based on a common dataset, indexed using the hypergraph-of-entity. A single index is used to evaluate three different tasks, based on a universal ranking function and a set of three separate relevance judgments.

The remainder of this article is organized as follows. In Section~\ref{sec:ref}, we present a short overview on keyword extraction, identifying graph-based approaches. This is then followed by an introduction to entity-oriented search, surveying approaches applied to each of the four main tasks we identify. In Section~\ref{sec:meth} we present the materials and methods that support this contribution. Particularly, in Section~\ref{sec:meth:inex}, we introduce the INEX 2009 Wikipedia collection, explaining why use a \mbox{10-year-old} collection instead of a more recent alternative, as well as highlighting the limitations of working with encyclopedic data. In Section~\ref{sec:meth:kw}, we describe the TextRank simplification we used to build document profiles of a shorter length. We also provide a study of the influence of the cutoff ratio applied to the selection of the top keywords. In Section~\ref{sec:meth:hgoe}, we present an overview on the hypergraph-of-entity, describing how the data structure is the main agent of ranking, making it possible for a universal ranking function to exist. In Section~\ref{sec:eval}, we present the results, describing the experimental framework, and presenting and commenting on the performance of each task over the joint representation model and the universal ranking function. Finally, in Section~\ref{sec:concl}, we present the conclusions, situating general retrieval models, such as the hypergraph-of-entity, in the state of the art, and proposing several lines of future work.

\section{Related Work}
\label{sec:ref}

We begin this section by presenting an overview on popular keyword extraction techniques, contextualizing their application to the hypergraph-of-entity and its reduction in size. We then introduce entity-oriented search and its main tasks, identifying commonalities that show how the hypergraph-of-entity can be used as a general entity-oriented search model.

\subsection{Efficient Keyword Extraction for Building Reduced Document Profiles in Graph-Based Scenarios}
\label{sec:ref:kw}

Keyword extraction is the process of identifying significant or descriptive words or short expressions to illustrate a given document. In the hypergraph-of-entity, both words and entities are represented as nodes, while relations are represented as hyperedges. The number of words that exist in a language is finite and the number of entities can also be considered finite, particularly for a given snapshot of a knowledge base, which is often used. This means that, as the index grows, the number of nodes will eventually converge. One way to reduce the size of the hypergraph-of-entity is to limit the number of nodes it contains. We experimented with keyword extraction to reduce document length and therefore the number of term nodes in the model. In this section, we explore several keyword extraction approaches, distinguishing between non-graph-based~\citep{Luhn1957,SparckJones1972,Campos2018} and graph-based~\citep{Mihalcea2004,Wan2008,Rose2010,Rousseau2015}.

\begin{table}[t]
    \centering
    \caption{Chronological overview of keyword extraction algorithms, identifying graph-based (GB) approaches.}
    \label{tab:ref:kw}
    \small

    \begin{tabularx}{\linewidth}{lRCp{30em}}
        \toprule
        \textbf{Algorithm} & \textbf{Year(s)} & \textbf{GB} & \textbf{Description}\\
        \midrule
        TF-IDF & \mbox{\citeyear{Luhn1957,SparckJones1972}} & \xmark & Document keywords selected based on whether they are frequent in the document, but rare in the collection.\\\\[-.5em]

        TextRank & \citeyear{Mihalcea2004} & \cmark & An undirected graph is built based on the co-occurrence of terms, optionally filtered by POS tags, within a sliding window. PageRank is applied, terms are ranked and adjacent terms form multi-term keywords.\\\\[-.5em]

        SingleRank & \citeyear{Wan2008} & \cmark & The input document is expanded with $k$ other similar documents. An undirected graph is built to link syntactically filtered words above a given affinity (weighted average of term frequencies per document similarity) threshold. Saliency is then computed based on the PageRank of this graph and adjacent top terms are merged into multi-term keywords.\\\\[-.5em]

        RAKE & \citeyear{Rose2010} & \cmark & Candidate keywords are generated by splitting the document by stopwords. A weighted graph of keyword co-occurrence is used to compute a keyword score (term frequency, degree and tf-degree ratio were tested as weighting functions). Wrongly split keywords are merged based on whether that instance repeats in the document and their scores are summed.\\\\[-.5em]

        Graph-of-Word & \citeyear{Rousseau2015} & \cmark & A directed graph is built based on a sliding window, filtered by noun and adjective POS tags. Keywords are then selected from the main core (i.e., largest the $k$-core). A $k$-core is a subgraph where every node has degree at least $k$. This way, there is no need to define a cutoff based on a threshold or ratio.\\\\[-.5em]

        YAKE! & \citeyear{Campos2018} & \xmark & Several term weighting functions are proposed and combined to score candidate keywords: casing, word position, word frequency, word relatedness to context, and word DifSentence. Multi-word keywords are then considered through the combination of candidate keyword scores, eliminating similar candidates through the Levenshtein distance.\\
        \bottomrule
    \end{tabularx}
\end{table}

Table~\ref{tab:ref:kw} provides a chronological overview of keyword extraction approaches. As we can see, many approaches are based on a graph of terms, which are frequently filtered by part-of-speech (POS) tags, in particular retaining nouns and adjectives. In TextRank~\citep{Mihalcea2004}, they used and undirected graph; in SingleRank~\citep{Wan2008}, they also used an undirected graph, but they also included terms from similar documents; in RAKE~\citep{Rose2010}, they used an undirected weighted graph; and in graph-of-word~\citep{Rousseau2015}, they used a directed graph. TextRank and SingleRank both relied on PageRank, while RAKE experimented with degree, and graph-of-word with maximal $k$-core retention. TextRank, SingleRank and RAKE also considered a post-processing stage, where keywords were merged into multi-term keywords. Out of the two non-graph-based approaches we covered, TF-IDF serves to illustrate one of the first, most iconic metrics of term importance, while YAKE!\ shows a state-of-the-art approach, based on other features that are not easily represented as a graph. In particular, these include: casing (ratio of uppercase term frequency to term frequency), word position (based on the median position of sentences containing the word), word frequency (divided by the sum of the mean and standard deviation), word relatedness to context (measuring the diversity of words co-occurring within a left and right window), and word DifSentence (the normalized number of sentences containing the word).

For the hypergraph-of-entity, the best keyword extraction method is not the most effective, but the most efficient. When working with general models, we must also consider whether the approach fits our current framework, as to prepare for a future integration leading to improved generality. Taking this into account means that, for our retrieval model, random walk and graph-based approaches are preferred. This leaves both TextRank and SingleRank as ideal candidates, since they both rely on graphs and PageRank, a random walk based approach. We selected TextRank, as SingleRank would expand to similar documents too prematurely for our model. This would have represented not only additional overhead, but also a redundant step that would have been analogously taken by the random walk score in hypergraph-of-entity during search. In Section~\ref{sec:meth:kw}, we present the details on how we further modified TextRank to reduce computation time, with very little impact in effectiveness for our retrieval model.

\subsection{Entity-Oriented Search: Semantic Retrieval over Corpora and Knowledge Bases}
\label{sec:ref:eos}

\citet[§1.3]{Balog2018} defined entity-oriented search as \textit{``the search paradigm of organizing and accessing information centered around entities, and their attributes and relationships''}. While many tasks can be designed around this definition, we focus on the following four main tasks: ad hoc document retrieval, ad hoc entity retrieval, related entity finding, and entity list completion.

In this context, only ad hoc document retrieval approaches, leveraging entities, are considered to be a part of this task. Some authors have also described the ranking of documents through an entity-informed approach as `semantic search'~\citep{Mangold2007}. On the other hand, `semantic search' has been used to describe ad~hoc entity retrieval, usually when referring to entity ranking over an RDF store~\citep{Blanco2011a}. \citet[§1.3.3]{Balog2018} has generalized the definition by stating that \textit{``Semantic search encompasses a variety of methods and approaches aimed at aiding users in their information access and consumption activities, by understanding their context and intent''}. Across different evaluation forums, there are also inconsistent designations and problem definitions for similar tasks. For example, `entity list completion' was a task named in TREC Entity track, but the INEX Entity Ranking track used the shorter name `list completion' for a similar task. In the TREC version, the input was an entity, a target type and a set of example entities~\citep[§3]{Balog2010}. In the INEX version, the input was a textual description of the context, as well as a set of example entities~\citep[§2.2]{Demartini2009}. The main goal in both tasks was, however, to complete a list of entities by ranking similar entities. This fits the more general and simplified definition that we adopt for this work and that we described in Section~\ref{sec:intr}.

While most of the definitions in entity-oriented search are only recently converging, we can still identify and classify previous scientific work within each, or even multiple, of the four identified tasks. Ad hoc document retrieval includes for instance the work by \citet{Bendersky2012} on the query hypergraph, where they modeled the dependencies between query concepts (e.g., terms, bigrams, named entities) and the document, as a log-linear combination of factors. It also includes the work by \citet{Devezas2019}, where they assessed the performance of the hypergraph-of-entity for the task of ad hoc document retrieval, based on a random walk ranking function. \citet{Bhagdev2008} have combined an inverted index and a triplestore to provide both ad hoc document, and entity retrieval, through a document URI identifier used to establish a relation of provenance to the entities in the knowledge base. \citet{Bast2013} have also explored the same two tasks by proposing a joint index for ontologies and text.

\citet{Zhiltsov2013} captured the latent semantics of entity-relations based on tensor factorization, considering term based features, as well as structural features. This supported the retrieval of entities based on keyword queries. \citet{Zhiltsov2015} proposed the fielded sequential dependence model (FSDM), an expansion of SDM by \citet{Metzler2005} that accounts for document field weights. They tackled ad hoc entity retrieval by defining entity profiles based on virtual documents, each containing the names, attributes, and categories of the indexed entity, as well as the names of similar and related entities. Later on, \citet{Nikolaev2016} expanded on this idea proposing two parameterized versions of FSDM, based on sequential and full term dependencies. Recently, \citet{Dietz2019} proposed ENT Rank (entity-neighbor-text), a hypergraph-based approach for entity ranking, where text was used to inform and improve entity retrieval. The hypergraph was converted into an entity co-occurrence multigraph and several features were considered to train a learning-to-rank-entities model: neighbor features, relation-typed neighbor features, and context-relevance features.

\citet{Zhou2014} has explored ad hoc document, and entity retrieval, as well as related entity finding, by focusing on the concept of querying by entities and/or for entities, varying the input and output as any combination of text and entities (e.g., in querying by entities, entities are taken as input, and documents are ranked as output). This is similar to the approach of the hypergraph-of-entity~\citep{Devezas2019}, where, by restricting the type of input (term or entities nodes, or document hyperedges), and output (i.e., collecting only terms, entities or documents), several tasks could be, in theory, generalized. \citet{Bron2013} explored entity list completion using text-based and structure-based approaches, as well as a combination of both, finding unstructured and structured data to benefit from each other in improving performance. The hypergraph-of-entity is a joint representation of text, entities and their relations that builds on this same idea. In this work we assess whether these four tasks are able to perform appropriately in such a generalized retrieval model.

\section{Materials and Methods}
\label{sec:meth}

In this section, we begin by describing the INEX 2009 Wikipedia collection, the dataset that we relied upon for evaluation [§\ref{sec:meth:inex}]. We then detail the simplification of the keyword extraction TextRank algorithm, the construction of the document profiles, and the selection of the appropriate ratio parameter [§\ref{sec:meth:kw}]. Finally, we present an overview of the hypergraph-of-entity, the proposed general representation and retrieval model, along with its universal ranking function [§\ref{sec:meth:hgoe}].

\subsection{Dataset: INEX 2009 Wikipedia Collection}
\label{sec:meth:inex}

The INEX 2009 Wikipedia collection~\citep{Schenkel2007} is a snapshot of the English Wikipedia, from October 8, 2008. It contains over $2.6$ million articles in XML format with over $100$ million XML elements. Documents were annotated with semantic information from the 2008-w40-2 version of the YAGO ontology, assigning labels based on one or multiple of the $5,800$ available classes (e.g., person, movie, city). This resulted in a dataset that uses $50.7$ GB of space when uncompressed, and $5.5$ GB when compressed, in four gzipped tar archives of $1.4$ GB each.

\subsubsection{Why use such an outdated collection?}

Despite being a 10-year-old test collection, INEX 2009 Wikipedia collection is still one of the few datasets that simultaneously provides relevance judgments for ad hoc document retrieval, ad hoc entity retrieval, and entity list completion. It is also adequate for experimenting with a combination of unstructured and structured data, since it contains both text, usually representing an entity, and links among these entities. Furthermore, while we do not take advantage of it in this work, the semantic annotations are also useful for identifying entity types, supporting for example type queries~\citep[§2.1]{Pound2010}. Besides this collection, another, more recent dataset that could be used for evaluating a general retrieval model would be TREC Washington Post Corpus\footnote{\url{https://trec.nist.gov/data/wapost/}}, since the tasks around it tackle both text and entity retrieval.

\subsubsection{Limitations of using an encyclopedic collection}

Knowledge bases like Wikipedia (semi-structured), DBpedia (structured), or Wikidata (structured) are frequently used to augment a corpus. A typical preprocessing pipeline in entity-oriented search is to annotate documents through named entity recognition and disambiguation, linking each entity to its entry in the knowledge base. This can then be used to improve document, or entity retrieval. However, when working with an encyclopedic test collection lik INEX 2009 Wikipedia collection, the corpus and the knowledge base are greatly self-contained. On one side, this means that it's harder to improve retrieval effectiveness through external knowledge, which is sparser. On the other side, it means that the collection doesn't need to be augmented, but rather preprocessed so that text, entities and their relations are extracted from the structure of the XML document.

\subsection{Building Document Profiles based on a TextRank Simplified for Efficiency}
\label{sec:meth:kw}

We use a simplified version of \mbox{TextRank} to build document profiles, based on the keywords extracted from each article in the INEX 2009 Wikipedia collection (Section~\ref{sec:meth:inex}). This way, we obtain shorter documents that are representative of the original Wikipedia articles, but require less space during indexing. We also study the behavior of the \textit{ratio} parameter, based on a smaller subset of the test collection, in order to determine the ideal fraction of keywords required for a good performance. Keyword extraction is done using a simplified version of TextRank~\citep{Mihalcea2004} over the preprocessed text (i.e., lower case, tokenized, without stopwords). We ignore POS tagging, syntactic filtering, and keyword collapse. Each pair of terms within a sliding window of size $n = 4$ is represented as two nodes connected in an undirected graph. PageRank is then computed for this graph, and term nodes are ranked accordingly. A fraction of the top keywords, defined by a \textit{ratio} parameter, is then used to represent the document in the index. This way, documents with a higher number of distinct terms will also be represented by a higher number of keywords.

\begin{table}[t]
    \centering
    \caption{Evaluating TF-IDF, BM25 and RWS performance for top keyword cutoffs between 1\% and 30\% (`--' for full-text). Index size is provided in megabytes (MiB), along with the number of nodes and hyperedges, when available. Effectiveness is measured using P@10, NDCG@10, MAP and GMAP. We show in bold the best values per ranking function, over keyword based runs (i.e., ignoring full-text).}
    \label{tab:eval:keyword-ratios}

    \begin{adjustbox}{width=\linewidth}
        \begin{tabularx}{1.15\linewidth}{XCRRRRRRRr}
            \toprule
            \textbf{Index} & \textbf{Ranking} & \textbf{Ratio} & \textbf{Size} & \textbf{Nodes} & \textbf{Edges} & \textbf{P@10} & \textbf{NDCG@10} & \textbf{MAP} & \textbf{GMAP}\\
            \midrule
            \multirow{12}{*}{Lucene} & \multirow{6}{*}{$\operatorname{TF-IDF}$}
            & 0.01      & \bf 1.2 MiB	& --	& --	& \bf 0.1500	& \bf 0.1674	& 0.0769	& 6.50e-06 \\
            & & 0.05      & 3.0 MiB	& --	& --	& 0.0800	& 0.0720	& 0.1172	& 1.00e-05 \\
            & & 0.10      & 5.3 MiB	& --	& --	& 0.0600	& 0.0477	& 0.1310	& 1.31e-05 \\
            & & 0.20      & 9.9 MiB	& --	& --	& 0.0400	& 0.0316	& \bf 0.1335	& 1.31e-05 \\
            & & 0.30      & 15 MiB	& --	& --	& 0.0500	& 0.0380	& 0.1306	& \bf 1.43e-05 \\
            & & --        & 104 MiB	& --	& --	& 0.2800	& 0.2667	& 0.2160	& 1.76e-01 \\
            \cmidrule{3-10}
            & \multirow{6}{*}{$\underset{\begin{subarray}{c}k_1 = 1.2,\\b = 0.75\end{subarray}}{\operatorname{BM25}}$}
            & 0.01      & \bf 1.2 MiB	& --	& --	& \bf 0.1500	& \bf 0.1674	& 0.0768	& 6.48E-06 \\
            & & 0.05      & 3.0 MiB	& --	& --	& 0.0900	& 0.0734	& 0.1169	& 9.99E-06 \\
            & & 0.10      & 5.3 MiB	& --	& --	& 0.0600	& 0.0472	& 0.1315	& 1.31E-05 \\
            & & 0.20      & 9.9 MiB	& --	& --	& 0.0400	& 0.0285	& \bf 0.1331	& 1.30E-05 \\
            & & 0.30      & 15 MiB	& --	& --	& 0.0400	& 0.0285	& 0.1304	& \bf 1.43E-05 \\
            & & --        & 104 MiB	& --	& --	& 0.4900	& 0.5479	& 0.3412	& 3.15E-01 \\
            \midrule
            \multirow{6}{*}{HGoE} & \multirow{6}{*}{$\underset{\begin{subarray}{c}\ell = 2, r = 1000,\\ exp. = false\end{subarray}}{\operatorname{RWS}}$}
            & 0.01      & \bf 93 MiB	& \bf 291k	& \bf 146k & 0.2900 & 0.2581 & 0.1542 & 9.22e-02 \\
            & & 0.05      & 101 MiB	& 301k	& 177k & 0.2700 & 0.2724 & \bf 0.1846 & \bf 1.07e-01 \\
            & & 0.10      & 106 MiB	& 312k	& 190k & \bf 0.3300 & \bf 0.3227 & 0.1845 & 1.06e-01 \\
            & & 0.20      & 114 MiB	& 334k	& 204k & 0.2600 & 0.2299 & 0.1559 & 9.09e-02 \\
            & & 0.30      & 122 MiB	& 360k	& 214k & 0.2400 & 0.2484 & 0.1534 & 9.36e-02 \\
            & & --        & 182 MiB	& 607k	& 253k & 0.1700 & 0.1671 & 0.1312 & 1.01e-01 \\
            \bottomrule
        \end{tabularx}
    \end{adjustbox}
\end{table}

\begin{figure}
  \subfigure[Storage efficiency.\label{fig:kw-cutoff:size}]{
    \includegraphics[width=.38\linewidth]{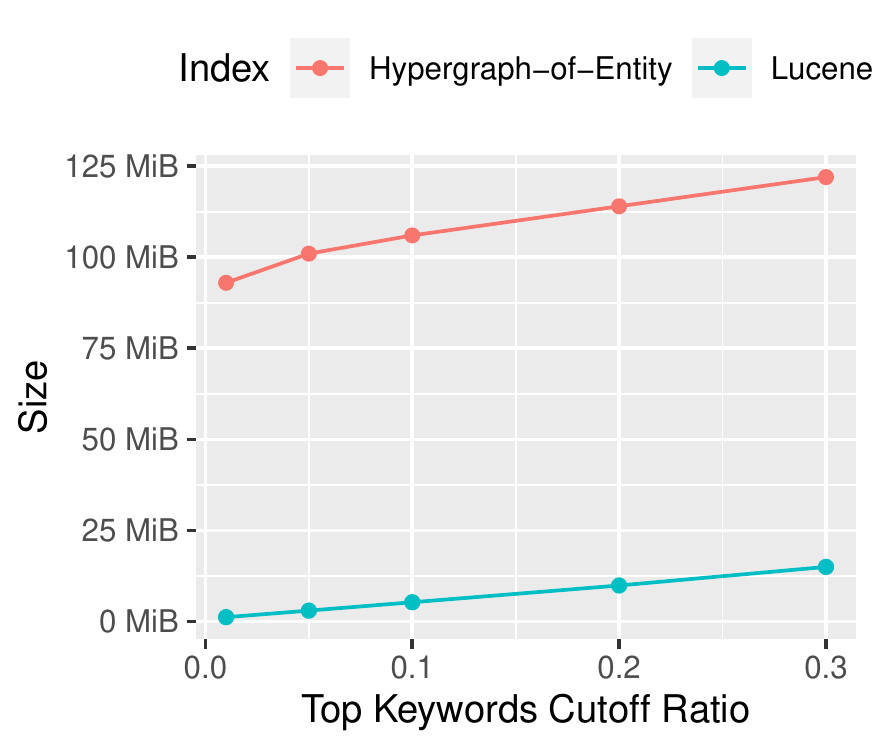}}
  \hfill
  \subfigure[Retrieval effectiveness.\label{fig:kw-cutoff:effectiveness}]{
    \includegraphics[width=.5975\linewidth]{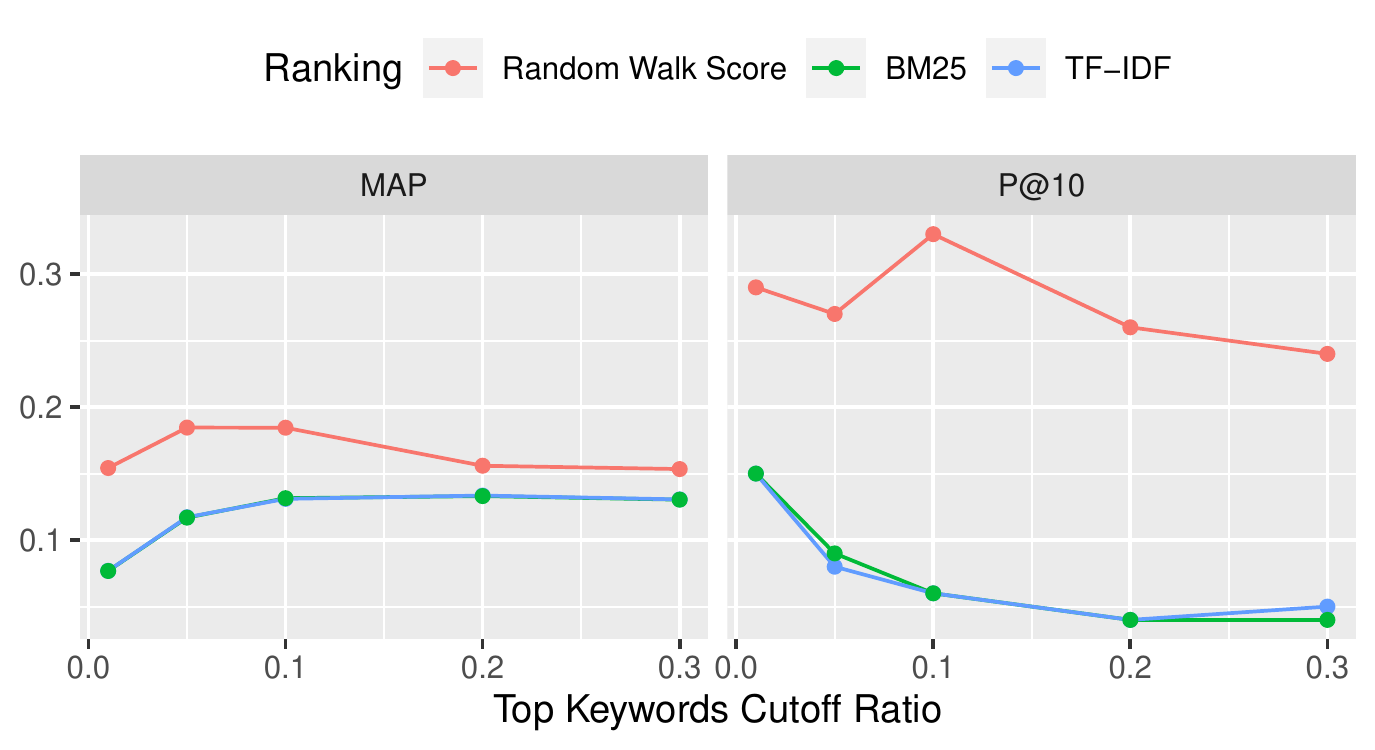}}
  \caption{Evolution of performance metrics for increasing cutoff ratios of top keywords.}
  \Description{Evolution of performance metrics for increasing cutoff ratios of top keywords.}
  \label{fig:kw-cutoff}
\end{figure}

In order to select the ideal fraction of keywords to use, we experimented with different ratio values based on a subset of the INEX 2009 Wikipedia collection and the evaluation data for the INEX~2010 Ad Hoc track. This subset, which we call INEX 2009 10T-NL, was built by randomly sampling 10~topics (`10T'), along with the documents mentioned in their relevance judgments, not including linked documents (`NL'). We compared the size of the generated index, in bytes as well as number of nodes and edges, for several ratio values: 0.01, 0.05, 0.10, 0.20, 0.30. We also assessed effectiveness for the ad hoc document retrieval task. For each run, we computed several performance metrics: P@10, NDCG@10, MAP, GMAP. As we can see in Table~\ref{tab:eval:keyword-ratios}, keyword extraction results in a considerable reduction, particularly for Lucene, where the index is $6.9 \times$ smaller for the top 30\% keywords. Similarly, the hypergraph-of-entity is reduced in size, resulting in a $1.5 \times$ smaller index for the top 30\% keywords. As we can also see in Figure~\ref{fig:kw-cutoff:size}, the size of the index decreases with the ratio, but always achieves a better reduction for Lucene. When considering the top 1\% keywords, the Lucene index is reduced $86.7 \times$, from 104~MiB to 1.2~MiB, while the hypergraph-of-entity is only reduced $2.0\times$, from 182~MiB to 93~MiB. In particular, the number of nodes is reduced $2.1\times$, from 607 to 291 thousand nodes, while the number of hyperedges is reduced $1.7\times$, from 253 to 146 thousand hyperedges. When considering the mean average precision (MAP), the complete indexes (i.e., with all document terms) achieve the best performance for Lucene with BM25. However, when considering any ratio, as shown in Figure~\ref{fig:kw-cutoff:effectiveness}, the version reduced to a document profile based on the top keywords consistently achieves a better MAP for the random walk score ($0.18$ for the top~1\% and top~5\% keywords). While the overall performance is lower, the random walk score is able to outperform TF-IDF and BM25, when available information is limited and, perhaps, more representative or selective. In fact, the lower the ratio, the better the MAP, reaching the ideal reduction and performance with the top 5\% keywords, after which MAP starts decreasing again, as there is too little information in the top~1\% keywords to discriminate the documents. Based on the results of this small-scale experiment, we opted for a ratio of $0.05$, despite $0.10$ having reached both a higher NDCG@10 and P@10 (cf.\ Figure~\ref{fig:kw-cutoff:effectiveness}). Our goal was to prioritize the minimization of space requirements.

\begin{figure}[t]
    \centering
    \includegraphics[width=\linewidth]{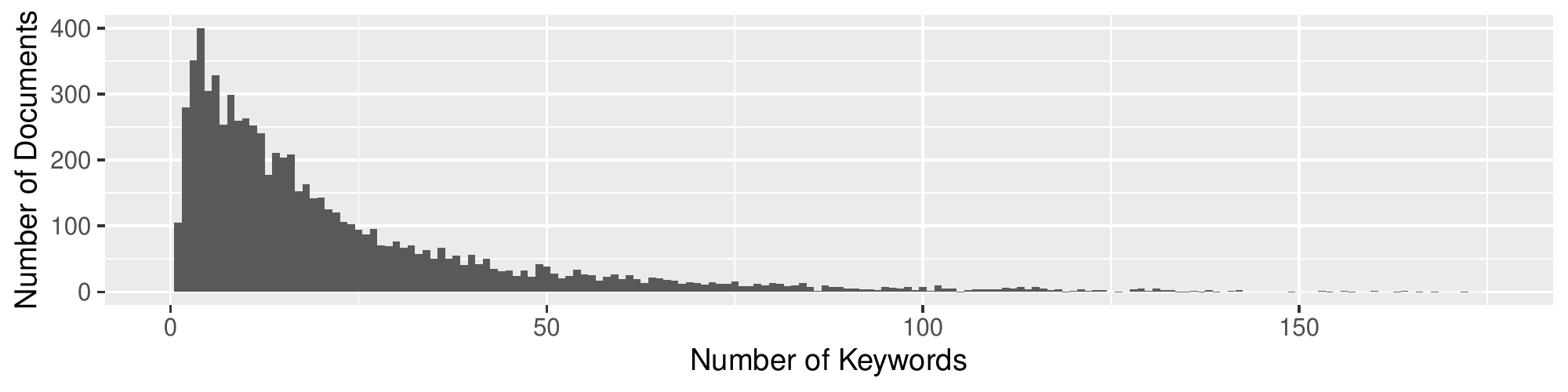}
    \caption{Keyword distribution for INEX 2009 10T-NL.}
    \label{fig:eval:kw_distr}
\end{figure}

Based on the simplified version of TextRank, the absolute number of keywords, retained based on the top~5\% per document, were distributed as illustrated in Figure~\ref{fig:eval:kw_distr}. On average, there were 22.9 keywords per document. There were 105 documents represented by only one keyword, 280 documents represented by two keywords, and 351 documents represented by three keywords. Most of the documents (400) were represented by four keywords, from then on following a logarithmic distribution, up to a maximum of 361 keywords (not displayed in the plot).

\subsection{Hypergraph-of-Entity: A General Representation and Retrieval Model for Entity-Oriented Search}
\label{sec:meth:hgoe}

The hypergraph-of-entity~\citep{Devezas2019} was proposed as a joint representation of terms, entities and their relations, for indexing corpora and knowledge bases in a unified manner. In this model, entities would link to other related entities as a group, either according to the knowledge base (e.g., the subject and the respective target objects), or simply the occurrence in a common document; terms and entities were all linked by a document hyperedge, as a bag of words and entities; and terms were linked to entities that they represented or illustrated in some way (e.g., based on string matching with the entity's name; perhaps also good for cross-language retrieval).

Several index extensions were also proposed, including synonyms, context, and TF-bins. Synonyms extended the model with terms from WordNet synsets, adding undirected hyperedges to group all synonym terms. Context was similar to synonyms, but instead of WordNet it relied on word2vec to obtain the most similar terms, also grouping them through an undirected hyperedge. TF-bins used undirected hyperedges to group terms within a given TF range for each document --- in this work we only use 2 TF-bins to distinguish between the most frequent and the least frequent terms per document.

The ranking function for this representation model was based on a series of repeated random walks of a fixed length, starting from a set of seed nodes representing the query. The proposed random walk score was able to cover nodes and hyperedges of all types, but only collected and ranked elements of the selected target type(s). The retrieval model is highly dependent of the representation model, as was well as the random walk score parameters.

In the following sections, we further detail aspects of the model relevant to this contribution. In Section~\ref{sec:meth:hgoe:repr} we discuss the representation model, describing how keyword extraction impacted the size of the model and highlighting the influence of the index structure in the ranking process. In Section~\ref{sec:meth:hgoe:rws}, we describe the random walk score, explaining how it can be used as a universal ranking function, greatly due to the representation-driven retrieval approach taken by the hypergraph-of-entity. We also describe the parameters we considered for this function in our experiments.

\subsubsection{A Representation-Driven Retrieval Indexing Structure}
\label{sec:meth:hgoe:repr}

The retrieval models based on indexing data structures like the inverted index are highly dependent on the ranking function that is chosen. These functions usually contain three main elements: $(1)$ term frequency, to measure the importance of query terms in the document; $(2)$ inverse document frequency, to diminish the impact of frequent terms that are widespread over the collection, therefore having little discriminative power; and $(3)$ pivoted document length normalization, to avoid long documents to outrank shorter, more relevant documents. While this is the standard, several other elements can be included based on an inverted index, such as term positions, term boosting payloads, or even document prior features that are query independent and stored in fields of the index. The final score, however, completely depends on the ranking function, that only uses the index to efficiently compute the statistics that it requires.

In the hypergraph-of-entity, however, the graph-based index data structure highly dictates the effectiveness of the ranking function. By reducing the number of possible paths linking terms and entities, as well as ensuring the quality of the retained terms, entities and relations, we increase the chances for the ranking function to succeed. A lower number of possible paths leads to lower uncertainty. In general, the lower the number of documents we consider, the lower the uncertainty as well. This is not to be confused with the quality of the proposed ranking, since smaller collections have a lower probability of having relevant documents for a wide range of topics. Lower uncertainty is related to the best possible answer the ranking function could return, based on the available information in the index. Given we cannot control the number of documents in a collection, which is often application-dependent, one way to decrease uncertainty is to rely on the reduction of the number of hyperedges and the improving of node and hyperedge quality. This is what keyword extraction does for the representation model in the hypergraph-of-entity. This is also why a lower number of representative keywords improves the effectiveness of the model, just until too little keywords are used (cf.\ Table~\ref{tab:eval:keyword-ratios}).

Keyword-based reduction directly impacted the number of nodes and indirectly impacted the number of hyperedges, since it reduced the number of links between terms and entities, due to a lower number of terms being considered. In our experiments, we used the base model of the hypergraph-of-entity, without synonymy or contextual similarity relations. This resulted in a hypergraph with $3,506,823$ nodes ($633,269$ terms and $2,873,554$ entities), as well as $7,721,743$ hyperedges ($2,653,452$ documents, $2,629,544$ entity relations using subject-based grouping, and $2,438,747$ text-entity relations based on the term occurrence in the entity name).

\subsubsection{The Random Walk Score as a Universal Ranking Function}
\label{sec:meth:hgoe:rws}

While the hypergraph-of-entity is a representation-driven retrieval model, the ranking function still requires configuration to answer each of the different entity-oriented search tasks. This is achieved both by controlling the type of input and output as described in Table~\ref{tab:hgoe:io}, and by configuring the parameters from Table~\ref{tab:hgoe:params}. The random walk score ranking function launches $r$ repeated random walks of a given length $\ell$, from each seed node. Seed nodes can either be the term nodes matching query terms, or their expansion to adjacent entity nodes. Each node and hyperedge has a counter that keeps track of the number of visits from the random walks, which are simulated step by step. This is then used to rank the desired output elements. As we can see in Table~\ref{tab:hgoe:io}, each of the four main tasks can be mapped to an input/output configuration, thus providing universal ranking for entity-oriented search. For example, by selecting a specific entity as the query (input), and ranking other entities (output), we are able to run the task of related entity finding.

\begin{table}[t]
    \centering
    \caption{Mapping entity-oriented search tasks to the hypergraph-of-entity.}
    \label{tab:hgoe:io}
    \small

    \begin{tabularx}{\linewidth}{lXlXl}
        \toprule
        & \textbf{Query} & \textbf{Input} & \textbf{Results} & \textbf{Output}\\
        \midrule
        \bf Ad hoc document retrieval& Keyword   & Term nodes            & Documents & Hyperedge ranking\\
        \bf Ad hoc entity retrieval  & Keyword   & Term nodes            & Entities  & Node ranking\\
        \bf Related entity finding   & Entity    & One entity node       & Entities  & Node ranking\\
        \bf Entity list completion   & Entity    & Multiple entity nodes & Entities  & Node ranking\\
        \bottomrule
    \end{tabularx}
\end{table}

\begin{table}[t]
    \centering
    \caption{Random walk score parameters and chosen configuration.}
    \label{tab:hgoe:params}
    \small

    \begin{tabularx}{\linewidth}{XlR}
        \toprule
        \textbf{Parameter} & \textbf{Description} & \textbf{Configuration}\\
        \midrule
        $\ell$                  & Length of the random walk.                        & $2$ \\
        $r$                     & Number of repeated random walks per seed node.    & $10,000$ \\
        $\mathid{expansion}$    & Whether to expand query to neighboring entities.  & $\mathid{false}$ \\
        $\mathid{directed}$     & Whether to consider or ignore direction.          & $\mathid{true}$ \\
        $\mathid{weighted}$     & Whether to consider node and hyperedge weights.   & $\mathid{false}$ \\
        \bottomrule
    \end{tabularx}
\end{table}

\section{Results}
\label{sec:eval}

While the hypergraph-of-entity is conceptually able to support multiple tasks, their individual performance, over a common index, still needs to be tested. In order to better understand the viability of the hypergraph-of-entity as a general model for entity-oriented search, we assess the effectiveness and efficiency for the following three tasks: $(1)$ ad hoc document retrieval (based on topics and qrels from INEX~2010 Ad Hoc track~\citep{Arvola2010}); $(2)$ ad hoc entity retrieval (based on topics and qrels from INEX~2009 Entity Ranking track~\citep{Demartini2009}, for the entity ranking task); and $(3)$ entity list completion (based on topics and qrels from INEX 2009 Entity Ranking track, for the list completion task). The approaches we describe here, including baselines, were developed as a part of Army ANT~\citep{Devezas2020}. This framework is available as open source software\footnote{Army ANT is available at: \url{https://github.com/feup-infolab/army-ant/tree/develop}.} and it can be used to reproduce these experiments\footnote{Please consult our YouTube videos to learn how to setup the framework: \url{https://www.youtube.com/playlist?list=PLc6NtbG0dqo1wGoYdTZkVd7I4SFNodKot}.}. The runs for each task were issued according to Table~\ref{tab:hgoe:io}, using the parameter configuration from Table~\ref{tab:hgoe:params} --- i.e., $\ell = 2$, $r = 10^4$, $\Delta_{nf} = 0$, $\Delta_{ef} = 0$, $\mathid{exp.} = F$, $\mathid{dir.} = T$, and $\mathid{wei.} = F$ --- except for TF-bins, where $\mathid{wei.} = T$ was used, assigning a default weight of $0.5$ to otherwise unweighted nodes and hyperedges.

Runs were based on the hypergraph-of-entity, built over the top 5\% keywords, extracted from each document of the complete INEX 2009 Wikipedia collection (Section~\ref{sec:meth:inex}), through TextRank. We also considered different versions of the hypergraph-of-entity: the \phdname{Base Model}, without index extensions; the \phdname{Syns} model, which added \edgetype{synonym} hyperedges, per term, based on WordNet synsets; the \phdname{Context} model, which added \edgetype{context} hyperedges, per term, based on the most similar other terms according to word2vec embeddings; the \phdname{TF-bins$_2$} model, which added \edgetype{tf\_bin} hyperedges for the most and least frequent terms per document; and the \phdname{Syns+Cont.} model, which included the hyperedges from the \phdname{Syns} and \phdname{Context} models.

Indexing took between $33h05m$ (\phdname{Syns}) and $37h16m$ (\phdname{Syns+Cont.}) to index on a virtual machine with a 4-core CPU and 32 GB of RAM. The baselines were supported on two Lucene indexes. The first was based on a text-only representation, using the extracted keywords, and it only took $15h06m$ to index. The second was based on an entity profile, built on the keywords extracted from a virtual document created from the concatenation of sentences mentioning the entity, which took $59h17m$ to index. This approach has been documented in the literature, notably in the work by \citet{Bautin2007}. We call these two Lucene indexes \phdname{Document Index} and \phdname{Entity Index}.

For the ad hoc entity retrieval task over the \phdname{Entity Index}, entities were ranked based on the keyword query issued over the Lucene virtual document. For the entity list completion over the \phdname{Entity Index}, entities were ranked based on an entity query, by first retrieving the virtual document for each entity in the query, and building a concatenated entity profile. This was then used to issue a \mbox{\command{MoreLikeThis} query} over the Lucene index, based on the concatenated entity profiles that were loaded through a \command{StringReader}. We did not impose a minimum term or document frequency, since keywords do not repeat and should, by definition, result in a low document frequency (and thus a high IDF). We also relied on the default value of 25 top keywords according to TF-IDF (or, in practice, IDF, since $TF = 1$) to build the query responsible for retrieving the ranked entities with a similar profile to the input entity set.

\begin{table}[t]
    \centering
    \caption{Evaluation results for hypergraph-of-entity as a general retrieval model (best scores per task in bold).}
    \label{tab:eval}

    \begin{adjustbox}{width=\linewidth}
        \begin{tabularx}{1.3\linewidth}{llXRRRRY{4.5em}}
            \toprule
            \textbf{Index} & \textbf{Task} & \textbf{Ranking} & \textbf{Avg./query} & \textbf{MAP} & \textbf{GMAP} & \textbf{P@10} & \textbf{NDCG@10} \\
            \midrule

            \multicolumn{8}{c}{\textbf{Lucene}}\\

            \midrule

            \multirow{2}{*}{Doc. Index} & \multirow{2}{*}{Ad hoc document retrieval}
            & TF-IDF  & 460ms	& 0.0228 & 0.0000 & 0.0692 & 0.0778 \\
            & & BM25    & 370ms	& 0.0324 & 0.0000 & 0.1173 & 0.1274 \\
            \midrule
            \multirow{4.5}{*}{Ent. Index} & \multirow{2}{*}{Ad hoc entity retrieval}
            & TF-IDF  & 1s~370ms & 0.0373 & 0.0000 & 0.0636 & 0.0670 \\
            & & BM25    &   798ms & 0.0668 & 0.0000 & 0.1182 & 0.1165 \\
            \cmidrule{2-8}
            & \multirow{2}{*}{Entity list completion}
            & TF-IDF  & 1s~230ms & 0.0558 & 0.0044 & 0.1000 & 0.1014 \\
            & & BM25    & 1s~221ms & 0.0666 & 0.0067 & \textbf{0.1250} & \textbf{0.1212} \\

            \midrule

            \multicolumn{8}{c}{\textbf{Hypergraph-of-Entity}}\\

            \midrule

            \multirow{3}{*}{Base Model} & Ad hoc document retrieval & \multirow{3}{*}{RWS}
            & 23s~405ms & 0.0863 & 0.0278  & 0.2462 & 0.2662 \\
            & Ad hoc entity retrieval     &
            & 26s~330ms & \textbf{0.1390} & 0.0002  & 0.2455 & 0.2425 \\
            & Entity list completion      &
            & 19s~162ms & 0.0879 & \textbf{0.0376}  & 0.0769 & 0.0594 \\

            \midrule

            \multirow{3}{*}{Syns} & Ad hoc document retrieval & \multirow{3}{*}{RWS}
            & 55s~555ms & \textbf{0.0937}  & \textbf{0.0303} & 0.2615 & 0.2812 \\
            & Ad hoc entity retrieval     &
            & 30s~232ms & 0.1337 & \textbf{0.0004} & 0.2473 & \textbf{0.2445} \\
            & Entity list completion      &
            & 19s~875ms & 0.0857 & 0.0368  & 0.0635 & 0.0474 \\

            \midrule

            \multirow{3}{*}{Context} & Ad hoc document retrieval & \multirow{3}{*}{RWS}
            & 24s~348ms & 0.0869 & 0.0245  & 0.2654 & 0.2784 \\
            & Ad hoc entity retrieval     &
            & 27s~620ms & 0.1304 & 0.0002  & 0.2364 & 0.2298 \\
            & Entity list completion      &
            & 19s~422ms & 0.0875 & 0.0373  & 0.0692 & 0.0520 \\

            \midrule

            \multirow{3}{*}{TF-Bins$_2$} & Ad hoc document retrieval & \multirow{3}{*}{RWS}
            & 2m~58s & 0.0172 & 0.0033 & 0.0500 & 0.0508 \\
            & Ad hoc entity retrieval     &
            & 4m~41s & 0.0300 & 0.0000 & 0.1145 & 0.1307 \\
            & Entity list completion      &
            & 1m~08s  & 0.0006 & 0.0000 & 0.0058 & 0.0053 \\

            \midrule

            \multirow{3}{*}{Syns+Cont.} & Ad hoc document retrieval & \multirow{3}{*}{RWS}
            & 23s~265ms & 0.0882 & 0.0246 & \textbf{0.2692} & \textbf{0.28830} \\
            & Ad hoc entity retrieval     &
            & 26s~877ms & 0.1313 & 0.0002 & \textbf{0.2509} & 0.2422 \\
            & Entity list completion      &
            & 19s~824ms & \textbf{0.0884} & 0.0369 & 0.0788 & 0.0594 \\
            \bottomrule
        \end{tabularx}
    \end{adjustbox}
\end{table}

Table~\ref{tab:eval} illustrates the performance of each of the three evaluated tasks, over two Lucene and five hypergraph-of-entity representation models. With generalization in mind, our goal is to understand whether the universal ranking function that we propose is able to adequately provide answers for all the tasks, over a common index. This means that we do not necessarily expect performance improvements, but rather an indication that our representation and retrieval model has the potential to be iterated over and improved as a general solution for entity-oriented search (and eventually retrieval).

We provide two ranking function baselines based on Lucene TF-IDF and BM25, which are directly comparable with the random walk score for the same task on the hypergraph-of-entity models. While different indexes and retrieval strategies are required for each task when using Lucene, a single index and ranking function is sufficient to support the three tasks in the hypergraph-of-entity. As we can see, when using document or entity profiles based on the top~5\% keywords, the hypergraph-of-entity is able to outperform both Lucene TF-IDF and BM25 in every effectiveness metric, except P@10 and NDCG@10 for the entity list completion task. However, our model is considerably less efficient than Lucene, taking on average from $19s162ms$ to $4m21s$ to answer a query, when compared to only milliseconds with Lucene.

Overall, evaluation scores are low, possibly due to the limitations introduced when considering only the top 5\% keywords, however these values for the random walk score are up to par with (and in fact outperform) TF-IDF and BM25 in the same conditions. We assessed the statistical significance of the best model per task when compared with the respective BM25 baseline. Given the non-normality of the average precision scores, we relied on the Wilcoxon signed-rank test. We found that that the \phdname{Syns} model was significantly better than \phdname{BM25} in ad hoc document retrieval ($p<0.05$). The \phdname{Base Model} was significantly better than \phdname{BM25} in ad hoc entity retrieval ($p<0.05$). The \phdname{Syns+Cont.} was significantly better than \phdname{BM25} in entity list completion ($p = 0.057$).

We further compared the hypergraph-of-entity MAP score, for each of the three tasks, with the MAP and xinfAP~\citep{Yilmaz2008} scores from the INEX 2010 Ad Hoc track, and the INEX 2009 XER track. Since our model frequently ranked quite below the state-of-the-art, in performance, for previous experiments, we compared it with the lowest entries from participants in INEX. In the INEX~2010 Ad Hoc track~\citep{Arvola2010}, the lowest MAP was $0.3177$, compared to $0.0937$ for our best model. In the INEX 2009 XML Entity Ranking track~\citep{Demartini2009}, the lowest xinfAP for entity ranking was $0.082$, compared to a MAP of $0.1390$ for our best model, and the lowest xinfAP for entity list completion was $0.100$, compared to a MAP of $0.0884$ for our best model.

Despite the performance improvement over the baselines, when supported on keyword-based document profiles, the hypergraph-of-entity is still not expected to outperform Lucene TF-IDF or BM25, when relying on full-text indexing. This was visible through the comparison with the INEX 2009 and 2010 participant runs, but also hinted by the results found in Table~\ref{tab:eval:keyword-ratios}. There, we had found the full-text versions of Lucene TF-IDF and BM25 to consistently outperform all keyword-based versions. The opposite behavior was found for the hypergraph-of-entity, with the full-text versions consistently underperforming when compared to the keyword-based versions. This also suggests that the hypergraph-of-entity might require selective pruning to reduce the universe of possible paths and better guide the random walker. Moreover, our study was focused on a single test collection. In order to better understand the retrieval performance and the generality of the model, assessments over different test collections and across multiple tasks is still required.

\section{Conclusions}
\label{sec:concl}

We have further studied the hypergraph-of-entity, assessing it as a general retrieval model for entity-oriented search. Our focus was on whether we would be able to adequately support three specific tasks: ad hoc document retrieval (leveraging entities), ad hoc entity retrieval, and entity list completion. We compared the effectiveness of each task, with the lowest ranking runs of participants in the INEX 2009 and 2010 respective tasks. We found an overall lower performance, that consistently scaled with state-of-the-art evaluation scores (i.e., tasks that had generally higher scores, also had higher scores when compared with the remaining tasks in our model). However, for all the tested tasks, and according to MAP, GMAP, NDCG@10 and P@10, we were able to outperform Lucene TF-IDF and BM25 when representing documents by their keyword-based profile, with the exception of the entity list completion task based on P@10 and NDCG@10. This showed the potential for an effective hypergraph-of-entity model capable of supporting retrieval generalization. Despite its low overall performance, we have demonstrated that a unified framework for entity-oriented search can be built, and we have opened several new opportunities for contribution in improving the performance of the hypergraph-of-entity, motivating the proposal of new hypergraph-based retrieval models, or even the exploration of novel general retrieval models, studying the advantage of this new approach.

\subsection{Future Work}

Besides performance, there are also other conceptual contributions that can be explored as future work. The current ranking approach is based on simulating individual steps of the random walk, but ideally this would be based on a Markov process over a matrix or tensor representation of the general mixed hypergraph that is the base for our model. We could then take advantage of the GPU for improving efficiency. However, there are still several challenges. Only recently has CERN been studying algebraic approaches for representing general hypergraphs, using adjacency tensors~\citep{Ouvrard2017,Ouvrard2018}. However, our hypergraph is mixed (i.e., containing both directed and undirected hyperedges), which has not yet been explored. Furthermore, given the recency of this work, there are still no widespread tools for working with these tensors.

Other experiments with the hypergraph-of-entity could focus on alternative approaches to reduce the complexity of the representation model, such as the document profiles based on keywords that we proposed here. Can we automatically identify nodes or hyperedges that can be removed? How would other keyword extraction algorithms work, or how would other ratio values or even a fixed cutoff value perform to select the top keywords? Regarding the random walk score, which is nondeterministic, could we find a parameter that always results in convergence, no matter the dataset? Or could we instead experiment with reranking algorithms, such as learning to rank approaches? Would caching be sufficient? How would we apply learning to rank to three different tasks? Would we have to train a model for each task, defying the purpose of a general model, or could we train a joint model for all tasks? Another interesting line of research to follow would be the automatic generation of additional information to complement the results, based on the modeled relations in the hypergraph. One example of this would be the inclusion of an explanation or additional context on why the results were retrieved --- for example, the most relevant paths from query or seed nodes leading to the result.

In order to further support generality, we would also need to conduct experiments over different test collections, while modeling other retrieval or even recommendation tasks through the hypergraph-of-entity. The ideal heterogeneous test collections for this endeavor are sparse, particularly those that also provide topics and relevance judgments for multiple tasks. Despite this challenge, there are other test collections that can be explored, even if only partially respecting these requirements. One such collection is the DBpedia-Entity v2~\citep{Hasibi2017}, which is regarded by the entity-oriented search community as a recent and robust benchmark. In the future, we will prepare an experiment based on this collection, which will position the hypergraph-of-entity's performance in regards to the state of the art, while also reinforcing its generality as a unified framework for information retrieval.

The challenges in joint representation models and universal ranking functions are immense. They pave the way for unified information retrieval, where a single model might support tasks like search and recommendation, but also subtasks like query expansion, and word sense and named entity disambiguation.

\section*{Funding}

José Devezas is supported by research grant PD/BD/128160/2016, provided by the Portuguese national funding agency for science, research and technology, Fundação para a Ciência e a Tecnologia (FCT), within the scope of Operational Program Human Capital (POCH), supported by the European Social Fund and by national funds from MCTES.

\bibliographystyle{ACM-Reference-Format}
\bibliography{ggen2020}

\end{document}